\documentclass[journal]{IEEEtran}
\usepackage{bbm}
\usepackage{caption}
\usepackage{subfigure}
\usepackage{graphicx}
\usepackage{latexsym}
\usepackage{diagbox}
\usepackage{changepage}
\usepackage[fleqn]{amsmath}
\usepackage{amsmath}
\usepackage{amsfonts}
\usepackage{indentfirst}
\usepackage{CJK}
\usepackage{indentfirst}
\usepackage[varg]{txfonts}
\usepackage{stfloats}
\usepackage{multirow}%
\usepackage{booktabs}
\usepackage{color,soul}
\usepackage{epstopdf}
\usepackage{float}
\usepackage{bm}
\usepackage{makecell}
\usepackage{array}

\usepackage{bm}
\usepackage{mathtools}
\usepackage{geometry}
\usepackage[linesnumbered,ruled,commentsnumbered,longend]{algorithm2e}
\usepackage[linesnumbered,ruled,commentsnumbered,longend]{algorithm2e}
\makeatletter

\newcommand{\Rmnum}[1]{\expandafter\@slowromancap\romannumeral #1@}

\makeatother
\makeatletter

\newcommand{\qed}{\nobreak \ifvmode \relax \else
	\ifdim\lastskip<1.5em \hskip-\lastskip
	\hskip1.5em plus0em minus0.5em \fi \nobreak
	\vrule height0.75em width0.5em depth0.25em\fi}

\SetAlgoLongEnd

\ifCLASSINFOpdf
\else
\fi

\begin{document}

\title{{The Far-/Near-Field Beam Squint and Solutions for THz Intelligent Reflecting Surface Communications}}
\author{Wanming Hao, \emph{Member, IEEE}, Xiaobei You, Fuhui Zhou, \emph{Senior Member, IEEE},  Zheng Chu, \emph{Member, IEEE}, Gangcan Sun, Pei Xiao, \emph{Senior Member, IEEE}
 \thanks{W. Hao, X. You and G. Sun are with the School of Information Engineering,  Zhengzhou University, Zhengzhou 450001, China. (e-mail: iewmhao@zzu.edu.cn, 202012172013018@gs.zzu.edu.cn, iegcsun@zzu.edu.cn)}
 \thanks{F. Zhou is with the College of Electronic and Information Engineering, Nanjing University of Aeronautics and Astronautics,
 	Nanjing, 210000, China. (email: zhoufuhui@ieee.org)}
 \thanks{Z. Chu and P. Xiao are 5GIC \& 6GIC, Institute for Communication Systems (ICS), University of Surrey, Guildford GU2 7XH, UK. (Email: Andrew.chuzheng7@gmail.com, p.xiao@surrey.ac.uk)}}

\maketitle
\begin{abstract}
Terahertz (THz) and intelligent reflecting surface (IRS) have been regarded as two promising technologies to improve the capacity and coverage for future 6G networks. Generally, IRS is usually equipped with large-scale elements when implemented at THz frequency. In this case, the near-field model and beam squint should be considered. Therefore, in this paper, we investigate the far-field and near-field beam squint problems in THz IRS communications for the first time. The far-field and near-field channel models are constructed based on the different electromagnetic radiation characteristics. Next, we first analyze the far-field beam squint and its effect for the beam gain based on the cascaded base station (BS)-IRS-user channel model, and then the near-field case is studied. To overcome the far-field and near-field beam squint effects, we propose to apply delay adjustable metasurface (DAM) to IRS, and develop a scheme of optimizing the reflecting phase shifts and time delays of IRS elements, which effectively eliminates the beam gain loss caused by beam squint. Finally, simulations are conducted to demonstrate the effectiveness of our proposed schemes in combating the near and far field beam squint.
\end{abstract}

\begin{IEEEkeywords}
Terahertz, intelligent reflecting surface, beam squint, delay adjustable metasurface.
\end{IEEEkeywords}

%
\IEEEpeerreviewmaketitle

\section{Introduction}
Terahertz (0.1$\sim$10 THz) communication has been recognized as a promising candidate technology for future B5G/6G networks owing to its ultra-wideband bandwidth and ultra-high transmission rate~\cite{1a},~\cite{2a}. The 2019 World Radio Communication Conference has proclaimed that the frequency band 275-450 GHz with a total of 137 GHz bandwidth will be used for land mobile service applications~\cite{3a}. However, the attenuation of THz signals is large, and they are also issues with diffraction and obstacles~\cite{4a}. To tackle those problems, intelligent reflecting surface (IRS) composed of a large number of energy-efficient passive elements can be applied to improve the strength of THz signals or generate new links by adjusting the reflecting elements~\cite{5a,6a,7a,8}, and thus IRS technique constitutes a viable solution for future THz communications.

Generally, the structure of IRS elements is relatively simple, and only the amplitude and phase of reflection signals can be adjusted, which makes that IRS elements are frequency-independent. As a result, there exists beam squint in IRS communications. In fact, there have been several research focusing on beam squint~\cite{8a}. For example, Chen \emph{et al}.~\cite{9a} proposed a novel phase shift design scheme for mitigating beam squint for the line-of-sight (LoS) and non-Los (NLoS) scenarios, respectively. Specifically, for the former, they derived the optimal phase shift for each subcarrier and common phase shift matrix for all subcarriers. For the latter, they proposed a mean channel covariance matrix (MCCM) scheme to maximize the achievable rate. Wan \emph{et al.}~\cite{10a} designed a new set of basis vectors to approximate the optimal precoder, where each new basis vector was devised to form a radiation pattern with a wide beam for overcoming beam squint effect. Although the above proposed phase optimization algorithms can reduce beam squint effects, the results are not remarkably effective. Later, several new antenna structures are developed by adding time delayers between radio frequency (RF) chains and phase shifters, which effectively solving beam squint effects. For example, Tan and Dai~\cite{11a} designed a delay-phase precoding structure to solve beam squint effects by converting frequency-independent beams into frequency-dependent beams, obtaining the near-optimal beam gain at all subcarriers. Gao and Wang~\cite{12a} proposed a wideband hybrid beamforming technique based on the true-time-delay lines and designed a new hardware implementation of analog combiner to achieve the near-optimal performance. Zhai \emph{et al.}~\cite{13a} designed an antenna architecture called THzPrism, where several true-time-delay lines were inserted among phase shifters to achieve frequency-based beam spreading.

In addition, it is well known that the electromagnetic radiation field is divided into far-field region and near-field region according to
the Fraunhofer distance~\cite{14a},~\cite{15a}. As IRS size increases, its near-field range will be expanded, and thus the phase of array steering vector is nonlinear to the antenna index~\cite{15a}. There have been several works related to the near-field model. Cui \emph{et al.}~\cite{16a} investigated the near-field channel estimation problem in extremely large-scale multiple-input-multiple-output (XL-MIMO) systems and proposed on-grid and off-grid near-field channel estimation schemes to achieve a better normalized mean square error (NMSE) performance. Later, they studied the near-field beam squint effect in THz communications, and discovered a new phenomenon that the beam energy at different frequencies focus on different locations~\cite{17a}. On this basis, they proposed a phase-delay focusing method to mitigate the near-field beam squint effect. Zhang \emph{et al.}~\cite{18a} investigated the performance of IRS-aided system in both the far-field and near-field by deriving accurate closed-form approximations of the ergodic capacity.

In summary, based on the above analysis, the beam squint and near-field model have been studied. However, the works in~\cite{9a}-\cite{13a} only consider the far-field beam squint without IRS, and the works in~\cite{16a}-\cite{18a} only consider either the near-field beam squint without IRS or near-filed IRS communications without beam squint. Generally, for THz IRS communications with large-scale array antennas, the near-field, far-filed and beam squint all should be considered. To our best knowledge, this has not yet been investigated to date.

To fill this gap, in this paper, we first analyze the far-field and near-field channel models and beam squints in THz IRS communications, respectively. Specifically, by theoretical derivation, the far-field beam squint makes that the beams  reflected by IRS at different subcarriers squint toward different directions, while the near-field beam squint makes that the beams reflected by IRS at different subcarriers aim at different locations and most of beams fail to focus on the user. Next, to deal with the far-field and near-field beam squint effects, we propose an effective scheme based on the delay adjustable metasurface (DAM)~\cite{19a} technique that can adjust the delays of signals reflected by different IRS elements to solve severe beam gain loss caused by beam squint. On this basis, we propose a scheme of jointly optimizing the reflection phase shifts of IRS elements and the time delay of DAMs in far-field and near-field, respectively, which makes the beams at all subcarriers focusing on the desired direction or location.

The rest of this paper is organized as follows. The far-field and near-field channel models of IRS-aided wideband THz communications are introduced, respectively, in Section \uppercase\expandafter{\romannumeral2}. The far-field and near-field beam squint effects are revealed, respectively, in Section \uppercase\expandafter{\romannumeral3}. In Section \uppercase\expandafter{\romannumeral4}, we describe our proposed schemes for overcoming the far-field and near-field beam squint effects, respectively. Section \uppercase\expandafter{\romannumeral5} presents the simulation results to verify our proposed schemes, and final conclusions are drawn in Section \uppercase\expandafter{\romannumeral6}.

\emph{Notations:}
We use the following notations throughout this paper. The scalars and vectors are denoted by italic letters $x$ and bold-face lower-case $\bf{x}$, respectively; $(\cdot)^T$ denotes the transpose and $\left\vert{\cdot}\right\vert$ stands for the absolute operator of a complex scalar. min$\{a,b\}$ and max$\{a,b\}$ represent the minimum and maximum from the real number $a$ to $b$, respectively.

\section{The Far-Field and Near-Field Channel Models}
We consider a basic IRS-aided wideband THz communication as shown in Fig. 1. In this paper, we mainly focus on investigating the beam squint caused by IRS, and thus the BS and user are all assumed equipped with single antenna. Meanwhile, we assume that there is no direct path between BS and user~\cite{6a}.
The transmission bandwith, carrier frequency and subcarrier index are denoted as $B$, $f_{c}$ and $M$, respectively. Thus, the $m$-th subcarrier frequency $f_{m}$ can be written as
\begin{eqnarray}
    f_{m}=f_{c}+\frac{B}{M}(m-1-\frac{M-1}{2}).
\end{eqnarray}

Suppose there are $L_{1}$ propagation paths between BS and IRS, and $L_{2}$ propagation paths between IRS and user. Let $u(t)$ and $v(t)$ denote the channel impulse response (CIR) spanning from BS to IRS and from IRS to user, respectively~\cite{20a}. Therefore, the CIR between BS and the $r$-th IRS element can be expressed as~\cite{21a}
\begin{eqnarray}
    u_{r}(t)=\sum\nolimits_{l_{1}=1}^{L_{1}}{\bar{\alpha}}_{l_{1}}e^{-j2\pi{f_{c}}\tau_{l1,r}^{\rm{BR}}}\delta(t-\tau_{l1,r}^{\rm{BR}}),
\end{eqnarray}
where ${\bar{\alpha}_{l_{1}}}$ is the equivalent baseband complex gain of the $l_{1}$-th path and $\tau_{l1,r}^{\rm{BR}}$ is the time delay of the $l_{1}$-th path between BS and the $r$-th IRS element, where $l_{1}\in\{{1,2,\cdots,L_{1}}\}$ and $r\in\{{1,2,\cdots,R}\}$. Similar to $u_{r}(t)$, the CIR between the $r$-th IRS element and user can be expressed as
\begin{eqnarray}	
    v_{r}(t)=\sum\nolimits_{l_{2}=1}^{L_{2}}{\bar{\alpha}}_{l_{2}}e^{-j2\pi{f_{c}}\tau_{l2,r}^{\rm{RU}}}\delta(t-\tau_{l2,r}^{\rm{RU}}),
\end{eqnarray}
where ${\bar{\alpha}_{l_{2}}}$ is the equivalent baseband complex gain of the $l_{2}$-th path and $\tau_{l2,r}^{\rm{RU}}$ is the time delay of the $l_{2}$-th path between the $r$-th IRS element and user, where $l_{2}\in\{{1,2,\cdots,L_{2}}\}$.

Denote $f_{r}(t)$ as the signal reflected by the $r$-th IRS element, we have
\begin{eqnarray}	
    f_{r}(t)=\theta_{r}u_{r}(t)\ast{s(t)}
            =\theta_{r}\sum\nolimits_{l_{1}=1}^{L_{1}}{\bar{\alpha}}_{l_{1}}e^{-j2\pi{f_{c}}\tau_{l1,r}^{\rm{BR}}}s(t-\tau_{l1,r}^{\rm{BR}}),
\end{eqnarray}
where $\pmb{\theta}=[\theta_{1},\theta_{2},\cdots,\theta_{R}]^T$ and denoting $\theta_{r}=\beta_{r}e^{j\phi_{r}}$ as the reflection coefficient of the $r$-th IRS element, where $\beta_{r}\in[0,1]$ and $\phi_{r}\in[0,2\pi)$ are the amplitude reflection coefficient and phase shift of the $r$-th IRS element, respectively. The default value of the amplitude reflection coefficient is set as $\beta_{r}=1, \forall_{r}$ for maximizing the signal power by IRS. Additionally, $s(t)$ is the signal transmitted by BS. Therefore, $\pmb{\theta}$ can be simplified as
 \begin{eqnarray}
\pmb{\theta}=[e^{j\phi_{1}},e^{j\phi_{2}},\cdots,e^{j\phi_{R}}]^T.
\end{eqnarray}

\begin{figure}[t]
	\begin{center}
		\includegraphics[width=10cm,height=5cm]{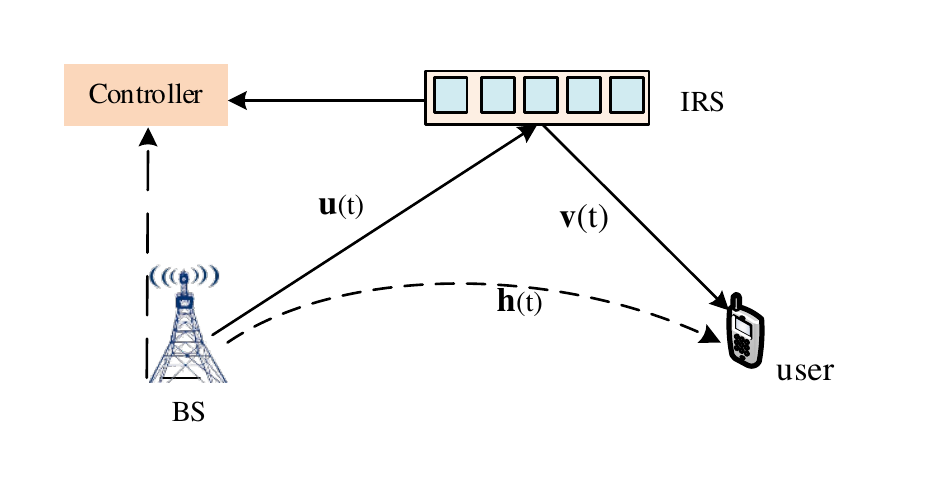}
		\caption{The system model of IRS-aided downlink wideband THz communications.}
		\label{figure1}
	\end{center}
\end{figure}

The $y_{r}(t)$ is the received signal of the user and denoted as
\begin{align}	
    y_{r}(t)=&v_{r}(t)\ast{f_{r}(t)}+n_{r}(t) \nonumber\\
    =&\sum\nolimits_{l_{2}=1}^{L_{2}}{\bar{\alpha}}_{l_{2}}e^{-j2\pi{f_{c}}\tau_{l2,r}^{\rm{RU}}}f_{r}(t-\tau_{l2,r}^{\rm{RU}})+n_{r}(t),
\end{align}
where $n_{r}(t)$ is the additive complex Gaussian noise. Substituting (4) into (6), $y_{r}(t)$ can be expressed as
\begin{align}	
    y_{r}(t)=&\theta_{r}\sum\nolimits_{l_{1}=1}^{L_{1}}\sum\nolimits_{l_{2}=1}^{L_{2}}{\bar{\alpha}}_{l_{1}}{\bar{\alpha}}_{l_{2}}
    e^{-j2\pi{f_{c}}\tau_{l1,r}^{\rm{BR}}}e^{-j2\pi{f_{c}}\tau_{l2,r}^{\rm{RU}}}\nonumber\\
    &\cdot{s(t-\tau_{l1,r}^{\rm{BR}}-\tau_{l2,r}^{\rm{RU}})}+n_{r}(t)\nonumber\\
    =&\theta_{r}h_{r}(t)\ast{s(t)}+n_{r}(t).
  \end{align}

According to (7), the CIR of the $r$-th cascaded BS-IRS-user channel can be expressed as
\begin{eqnarray}
    h_{r}(t)=\sum_{l_{1}=1}^{L_{1}}\sum_{l_{2}=1}^{L_{2}}{\bar{\alpha}}_{l_{1}}{\bar{\alpha}}_{l_{2}}e^{-j2\pi{f_{c}}\tau_{l1,r}^{\rm{BR}}}
    e^{-j2\pi{f_{c}}\tau_{l2,r}^{\rm{RU}}}\delta(t-\tau_{l1,r}^{\rm{BR}}-\tau_{l2,r}^{\rm{RU}}).
\end{eqnarray}

Based on the above, it is clear that the CIR of the $r$-th cascaded BS-IRS-user channel is relevant to the time delay $\tau_{l1,r}^{\rm{BR}}$ and $\tau_{l2,r}^{\rm{RU}}$. However, the time delay is different for the far-field and near-field. Next, based on the constructed CIR, we will analyze the far-field and near-field channel models, respectively.
\subsection{The Far-Field Channel Model}
In the far-field, it is usually assumed that the IRS array size is much smaller than the distance between BS and IRS as well as the distance between IRS and user. Based on this, the time delay $\tau_{l1,r}^{\rm{BR}}$ and $\tau_{l2,r}^{\rm{RU}}$ can be written as
\begin{eqnarray}	
    \tau_{l1,r}^{\rm{BR}}=\tau_{l1}^{\rm{BR}}+(r-1)\frac{d\sin\chi_{l_{1}}}{c}=\tau_{l1}^{\rm{BR}}+(r-1)\frac{\varphi_{l1}^{\rm{BR}}}{f_{c}},
\end{eqnarray}
\begin{eqnarray}	
    \tau_{l2,r}^{\rm{RU}}=\tau_{l2}^{\rm{RU}}-(r-1)\frac{d\sin\psi_{l_{2}}}{c}=\tau_{l2}^{\rm{RU}}-(r-1)\frac{\varphi_{l2}^{\rm{RU}}}{f_{c}},
\end{eqnarray}
where $c$ and $d$ are the speed of light and the IRS element spacing, respectively. We set $d$ as half wavelength, i.e., $d=\frac{\lambda_{c}}{2}$, where $\lambda_{c}$ is the carrier wavelength. For notational simplicity, we define $\tau_{l1}^{\rm{BR}}=\tau_{l1,1}^{\rm{BR}}$ and $\tau_{l2}^{\rm{RU}}=\tau_{l2,1}^{\rm{RU}}$.
Moreover, $\chi_{l_{1}}$ and $\psi_{l_{2}}$ represent the angle of arrival (AOA) of the $l_{1}$-th path between the BS and IRS as well as the angle of departure (AOD) of the $l_{2}$-th path between the IRS and user, respectively. Denote $\varphi_{l1}^{\rm{BR}}$ and $\varphi_{l2}^{\rm{RU}}$ as the normalized AOA and AOD with the range $[-\frac{1}{2},\frac{1}{2}]$, respectively.

According to (9) and (10), the CIR of the $r$-th cascaded BS-IRS-user channel in (8) can be obtained as
\begin{align}
    h_{r}^{\rm{far}}(t)=&\sum\nolimits_{l_{1}=1}^{L_{1}}\sum\nolimits_{l_{2}=1}^{L_{2}}{\bar{\alpha}}_{l_{1}}e^{-j2\pi{f_{c}}\tau_{l1}^{\rm{BR}}}
    {\bar{\alpha}}_{l_{2}}e^{-j2\pi{f_{c}}\tau_{l2}^{\rm{RU}}} \nonumber\\
    &\cdot{e^{-j2\pi{f_{c}}(r-1)\frac{\varphi_{l1}^{\rm{BR}}}{f_{c}}}
    e^{j2\pi{f_{c}}(r-1)\frac{\varphi_{l2}^{\rm{RU}}}{f_{c}}}\delta(t-\tau_{l1,r}^{\rm{BR}}-\tau_{l2,r}^{\rm{RU}})}.
\end{align}
Let $\alpha_{l_{1}}={\bar{\alpha}}_{l_{1}}e^{-j2\pi{f_{c}}\tau_{l1}^{\rm{BR}}}$ and $\alpha_{l_{2}}={\bar{\alpha}}_{l_{2}}e^{-j2\pi{f_{c}}\tau_{l2}^{\rm{RU}}}$, we have
\begin{align}
h_{r}^{\rm{far}}(t)=&\sum\nolimits_{l_{1}=1}^{L_{1}}\sum\nolimits_{l_{2}=1}^{L_{2}}\alpha_{l_{1}}\alpha_{l_{2}}e^{-j2\pi(r-1)\varphi_{l1}^{\rm{BR}}} \nonumber\\
    &\cdot{e^{j2\pi(r-1)\varphi_{l2}^{\rm{RU}}}\delta(t-\tau_{l1,r}^{\rm{BR}}-\tau_{l2,r}^{\rm{RU}})}.
\end{align}

By applying Fourier transform to (12), the frequency response corresponding of (12) can be expressed as
\begin{align}
h_{r}^{\rm{far}}(f)&=\int_{-\infty}^{+\infty}h_{r}^{\rm{far}}(t)e^{-j2\pi{f}t}\mathrm{d}t \nonumber\\
&=\sum_{l_{1}=1}^{L_{1}}\sum_{l_{2}=1}^{L_{2}}\alpha_{l_{1}}\alpha_{l_{2}}
e^{-j2\pi(r-1)(\varphi_{l1}^{\rm{BR}}-\varphi_{l2}^{\rm{RU}})}e^{-j2\pi{f}(\tau_{l1,r}^{\rm{BR}}+\tau_{l2,r}^{\rm{RU}})} \nonumber\\
&=\sum_{l_{3}=1}^{L_{1}L_{2}}\alpha_{l_{3}}^{C}e^{-j2\pi(r-1)\varphi_{l3}^{\rm{C}}(1+\frac{f}{f_{c}})}e^{-j2\pi{f}\tau_{l3}^{\rm{C}}},
\end{align}
where $\alpha_{l_{3}}^{C}$, $\varphi_{l3}^{\rm{C}}$ and $\tau_{l3}^{\rm{C}}$ are defined as the equivalent complex gain, angle and delay of the cascaded BS-IRS-user channel, respectively, where $\alpha_{l_{3}}^{C}=\alpha_{l_{1}}\alpha_{l_{2}}$, $\varphi_{l3}^{\rm{C}}=\varphi_{l1}^{\rm{BR}}-\varphi_{l2}^{\rm{RU}}$ and $\tau_{l3}^{\rm{C}}=\tau_{l1,r}^{\rm{BR}}+\tau_{l2,r}^{\rm{RU}}$. Additionally, we have $\varphi_{l3}^{\rm{C}}\in(-1,1)$ and $l_{3}\in{\{1, 2, \cdots, L_{1}L_{2}\}}$.
\subsection{The Near-Field Channel Model}
It is well known that as IRS array size increases, so does the Fraunhofer distance. Based on this, the CIR of the $r$-th cascaded BS-IRS-user channel can be written as
 \begin{align}
    h_{r}^{\rm{near}}(t)=&\sum\nolimits_{l_{1}=1}^{L_{1}}\sum\nolimits_{l_{2}=1}^{L_{2}}{\bar{\alpha}}_{l_{1}}{\bar{\alpha}}_{l_{2}}e^{-j2\pi{f_{c}}\tau_{l1,r}^{\rm{near,BR}}} \nonumber\\
    &e^{-j2\pi{f_{c}}\tau_{l2,r}^{\rm{near,RU}}}\delta(t-\tau_{l1,r}^{\rm{near,BR}}-\tau_{l2,r}^{\rm{near,RU}}),
\end{align}
where $\tau_{l1,r}^{\rm{near,BR}}$ denotes the time delay  of the $l_{1}$-th path between BS and the $r$-th IRS element, and $\tau_{l2,r}^{\rm{near,RU}}$ denotes the time delay of the $l_{2}$-th path between the $r$-th IRS element and the user, which can be defined as  $\tau_{l1,r}^{\rm{near,BR}}=\frac{d_{l1,r}^{\rm{BR}}}{c}$ and $\tau_{l2,r}^{\rm{near,RU}}=\frac{d_{l2,r}^{\rm{RU}}}{c}$,
where $d_{l1,r}^{\rm{BR}}$ is the distance of the $l_{1}$-th path from the BS to the $r$-th IRS element, and $d_{l2,r}^{\rm{RU}}$ is the distance of the $l_{2}$-th path from the $r$-th IRS element to the user.

Substituting $\tau_{l1,r}^{\rm{near,BR}}$ and $\tau_{l2,r}^{\rm{near,RU}}$ into (14), we have
\begin{align}
    h_{r}^{\rm{near}}(t)=&\sum\nolimits_{l_{1}=1}^{L_{1}}\sum\nolimits_{l_{2}=1}^{L_{2}}{\bar{\alpha}}_{l_{1}}e^{-j2\pi{f_{c}}\frac{d_{l1,r}^{\rm{BR}}}{c}} \nonumber\\
    &{\bar{\alpha}}_{l_{2}}e^{-j2\pi{f_{c}}\frac{d_{l2,r}^{\rm{RU}}}{c}}\delta(t-\frac{d_{l1,r}^{\rm{BR}}}{c}-\frac{d_{l2,r}^{\rm{RU}}}{c}).
\end{align}

Similar to (13), we take Fourier transform to (15), the frequency response of (15) can be expressed as
\begin{align}
h_{r}^{\rm{near}}(f)&=\int_{-\infty}^{+\infty}h_{r}^{\rm{near}}(t)e^{-j2\pi{f}t}\mathrm{d}t \nonumber\\
&=\sum_{l_{1}=1}^{L_{1}}\sum_{l_{2}=1}^{L_{2}}{\bar{\alpha}}_{l_{1}}{\bar{\alpha}}_{l_{2}}
e^{-j2\pi{f_{c}}\frac{d_{l1,r}^{\rm{BR}}}{c}}e^{-j2\pi{f_{c}}\frac{d_{l2,r}^{\rm{RU}}}{c}}e^{-j2\pi{f}(\frac{d_{l1,r}^{\rm{BR}}}{c}+\frac{d_{l2,r}^{\rm{RU}}}{c})} \nonumber\\
&=\sum_{l_{1}=1}^{L_{1}}\sum_{l_{2}=1}^{L_{2}}{\bar{\alpha}}_{l_{1}}{\bar{\alpha}}_{l_{2}}e^{-j2\pi{f_{c}}(\frac{d_{l1,r}^{\rm{BR}}}{c}+\frac{d_{l2,r}^{\rm{RU}}}{c})} e^{-j2\pi{f}(\frac{d_{l1,r}^{\rm{BR}}}{c}+\frac{d_{l2,r}^{\rm{RU}}}{c})} \nonumber\\
&=\sum_{l_{1}=1}^{L_{1}}\sum_{l_{2}=1}^{L_{2}}{\bar{\alpha}}_{l_{1}}{\bar{\alpha}}_{l_{2}}e^{-j\frac{2\pi}{\lambda_{c}}(1+\frac{f}{f_{c}})(d_{l1,r}^{\rm{BR}}+d_{l2,r}^{\rm{RU}})}.
\end{align}

It is clear that the phase $-\frac{2\pi}{\lambda_{c}}(1+\frac{f}{f_{c}})(d_{r}^{\rm{BR}}+d_{r}^{\rm{RU}})$ is non-linear to the IRS element index $r$ in the near-field, which is in contrast to the phase in the far-field, i.e., $-2\pi(r-1)\frac{d}{\lambda_{c}}(1+\frac{f}{f_{c}})\upsilon$ is linear to the IRS element index $r$.
\section{The Far-field and Near-Field Beam Squint Effects}
In this section, the far-field and near-field beam squint effects are analyzed, respectively. Specially, the far-field beam squint makes that the beams at different subcarriers squint toward different directions, which is distinct from that in the near-field, where the beam squint generates a new effect as the beams at different subcarriers spread to different locations.
\subsection{The Far-Field Beam squint Effect}

For analyze the far-field beam squint effect, we consider a single path and set as $L_{1}=1$, $L_{2}=1$  for sake of simplicity. Thus, the far-field channel can be rewritten as
\begin{eqnarray}
h_{r}^{\rm{far}}(f)=\alpha{e^{-j2\pi(r-1)\varphi(1+\frac{f}{f_{c}})}}e^{-j2\pi{f}\tau_{0}},
\end{eqnarray}
where $\varphi=\frac{d}{\lambda_{c}}(\sin\chi-\sin\psi)$ denotes the equivalent angle of the cascaded BS-IRS-user channel and $\tau_{0}=\tau_{0}^{\rm{BR}}+\tau_{0}^{\rm{RU}}$ is the equivalent time delay. Let $\upsilon=\sin\chi-\sin\psi$, we have
\begin{eqnarray}
h_{r}^{\rm{far}}(f)=\alpha{e^{-j2\pi(r-1)\frac{d}{\lambda_{c}}(1+\frac{f}{f_{c}})\upsilon}}e^{-j2\pi{f}\tau_{0}}.
\end{eqnarray}

Based on (18), the cascaded channel vector ${\bf{h}}(f)=[h_{1}(f),h_{2}(f),\cdots,h_{R}(f)]^{T}$ can be expressed as
\begin{eqnarray}
{\bf{h}}^{\rm{far}}(f)=\alpha{{\bf{a}}((1+\frac{f}{f_{c}})\varphi)}e^{-j2\pi{f}\tau_{0}},
\end{eqnarray}
where
\begin{eqnarray}
{\bf{a}}((1+\frac{f}{f_{c}})\varphi)=[1,e^{-j2\pi\frac{d}{\lambda_{c}}(1+\frac{f}{f_{c}})\upsilon},\cdots,
e^{-j2\pi(R-1)\frac{d}{\lambda_{c}}(1+\frac{f}{f_{c}})\upsilon}]^T, \nonumber
\end{eqnarray}
is the spatial steering vector. We define $g(f_{m},\nu,\phi_{r})=\left\vert{\bf{a}}^T((1+\frac{f_{m}}{f_{c}})\varphi)\pmb{\theta}^{\rm{far}}\right\vert$ as the gain at frequency $f_{m}$ and it can be expressed as
\begin{align}
g(f_{m},\nu,\phi_{r})=&\left\vert{\bf{a}}^T((1+\frac{f_{m}}{f_{c}})\varphi)\pmb{\theta}^{\rm{far}}\right\vert\nonumber\\
=&\left\vert \sum_{r=1}^{R}e^{-j2\pi(r-1)\frac{d}{\lambda_{c}}(1+\frac{f_{m}}{f_{c}})\nu}e^{j\phi_{r}}
\right\vert\nonumber\\
=&\left\vert \sum_{r=1}^{R}e^{j[\phi_{r}-\pi(r-1)(1+\frac{f_{m}}{f_{c}})\nu]} \right\vert.
\end{align}

By setting $f_{m}=f_{c}$ and $\varphi=\varphi_{0}$ with $\varphi_{0}=\frac{d}{\lambda_{c}}(\sin\chi_{0}-\sin\psi_{0})$, one can observe when $\phi_{r}-\pi(r-1)(1+\frac{f_{c}}{f_{c}})\nu_{0}=0$, (20) can reach its maximum value, and thus we have
\begin{eqnarray}
\phi_{r,c}=2\pi(r-1)\nu_{0}.
\end{eqnarray}

According to (21), the reflection phase shift vector $\pmb{\theta}_{c}^{\rm{far}}$ can be written as
\begin{eqnarray}
\pmb{\theta}_{c}^{\rm{far}}=[1,e^{j2\pi\nu_{0}},\cdots,e^{j2\pi(R-1)\nu_{0}}]^T.
\end{eqnarray}

Therefore, the beam gain achieved by the user under an arbitrary direction $\varphi$ at frequency $f_{m}$ can be denoted as
\begin{align}
g(f_{m},\nu,\phi_{r,c})=&\left\vert{\bf{a}}^T((1+\frac{f_{m}}{f_{c}})\varphi)\pmb{\theta}_{c}^{\rm{far}}\right\vert\nonumber\\
=&\left\vert \sum_{r=1}^{R}e^{-j2\pi(r-1)\frac{d}{\lambda_{c}}(1+\frac{f_{m}}{f_{c}})\nu}{e^{j2\pi(r-1)\nu_{0}}}
\right\vert\nonumber\\
=&\left\vert \sum_{r=1}^{R}e^{j\pi(r-1)[2\nu_{0}-(1+\frac{f_{m}}{f_{c}})\nu]} \right\vert.
\end{align}

To maximize the beam gain, it is observed that $2\nu_{0}-(1+\frac{f_{m}}{f_{c}})\nu=0$ should be satisfied. In this case, the beam direction can be expressed as

\begin{eqnarray}
\nu=\frac{2}{1+{f_{m}}/{f_{c}}}\nu_{0}.
\end{eqnarray}

From (24), one can observe $\nu\thickapprox\nu_{0}$ for any subcarrier in low frequency narrowband system due to $\frac{f_{m}}{f_{c}}\thickapprox1$. This means that the beams at all subcarriers will focus on the same direction and the maximum beam gain of each subcarrier can be obtained at this direction. However, for the ultra-wide band THz system, ${f_{m}}/{f_{c}}\thickapprox1$ no longer holds, specifically for edge subcarriers, it indicates that angle $\nu$ changes with the frequency $f_{m}$. This means that the beams at different subcarriers will squint toward different directions. Moreover, if we set ${\bf{a}}^T((1+\frac{f_{m}}{f_{c}})\varphi)={\bf{a}}^T((1+\frac{f_{m}}{f_{c}})\varphi_{0})$, for subcarrier frequency $f_{m}$, (20) should be
\begin{align}
g(f_{m},\nu_{0},\phi_{r,c})=&\left\vert{\bf{a}}^T((1+\frac{f_{m}}{f_{c}})\varphi_{0})\pmb{\theta}_{c}\right\vert\nonumber\\
=&\left\vert \sum_{r=1}^{R}e^{j\pi(r-1)[(1-\frac{f_{m}}{f_{c}})\nu_{0}]} \right\vert.
\end{align}

Similarly, it is clear that the beam gain $g(f_{m},\nu_{0},\phi_{r,c})$ reaches its maximum value since $f_{m}\thickapprox{f_{c}}$ in narrowband systems, while (25) can only obtain its maximum value around the central subcarriers in wideband THz IRS systems. This means that the beams at different subcarriers will be steered towards different directions. We will show the above phenomenon in simulation results section.

\subsection{The Near-Field Beam squint Effect}
 In the previous section, we have analyzed the characteristic of the beam squint in THz IRS communications based on far-field, and next we study the near-far beam squint.

 We assume that the location axis of BS and user are, respectively, $(0,0)$ and $(x,y)$, and $(x_{R},y_{R})$ denotes the axis of the first IRS element. Thus, the axis of the $r$-th IRS element is $(x_{R}+(r-1)d,y_{R})$, and we assume that IRS is an ($1\times{R}$) ULA for the convenience.
Here, we still consider a single path and set $L_{1}=1$, $L_{2}=1$ to ease explanation of the beam squint. Thus, the near-field channel can be expressed as
\begin{eqnarray}
h_{r}^{\rm{near}}(f)={\alpha}e^{-j\frac{2\pi}{\lambda_{c}}(1+\frac{f}{f_{c}})(d_{r}^{\rm{BR}}+d_{r}^{\rm{RU}})}.
\end{eqnarray}

According to (26), the cascaded channel vector ${\bf{h}}^{\rm{near}}(f)=[h_{1}^{\rm{near}}(f),h_{2}^{\rm{near}}(f),\cdots,h_{R}^{\rm{near}}(f)]^{T}$ can be written as
\begin{eqnarray}
{\bf{h}}^{\rm{near}}(f)=\alpha{{\bf{b}}((1+\frac{f}{f_{c}})(d_{r}^{\rm{BR}}+d_{r}^{\rm{RU}}))}e^{-j2\pi{f}\tau_{0}},
\end{eqnarray}
where
\begin{align}
{\bf{b}}((1+\frac{f}{f_{c}})(d_{r}^{\rm{BR}}+d_{r}^{\rm{RU}}))=&[1,e^{-j\frac{2\pi}{\lambda_{c}}(1+\frac{f}{f_{c}})(d_{1}^{\rm{BR}}+d_{1}^{\rm{RU}})},\cdots,\nonumber\\
&e^{-j\frac{2\pi}{\lambda_{c}}(1+\frac{f}{f_{c}})(d_{R}^{\rm{BR}}+d_{R}^{\rm{RU}})}]^T, \nonumber
\end{align}
is the spatial steering vector. Based on the definition of the beam gain in the previous section, for subcarrier frequency $f_{m}$ at location $(x',y')$, we have
\begin{align}
g(f_{m},d_{r}^{\rm{RU'}},\phi_{r}^{\rm{near}})=&\left\vert{\bf{b}}^T((1+\frac{f_{m}}{f_{c}})(d_{r}^{\rm{BR}}+d_{r}^{\rm{RU'}}))\pmb{\theta}^{\rm{near}}\right\vert\nonumber\\
=&\left\vert \sum_{r=1}^{R}e^{-j\frac{2\pi}{\lambda_{c}}(1+\frac{f_{m}}{f_{c}})(d_{r}^{\rm{BR}}+d_{r}^{\rm{RU'}})}e^{j\phi_{r}^{\rm{near}}} \right\vert \nonumber\\
=&\left\vert \sum_{r=1}^{R}e^{j[\phi_{r}^{\rm{near}}-\frac{2\pi}{\lambda_{c}}(1+\frac{f_{m}}{f_{c}})(d_{r}^{\rm{BR}}+d_{r}^{\rm{RU'}})]} \right\vert,
\end{align}
where $d_{r}^{\rm{RU'}}$ is the distance between the $r$-th IRS element and the location $(x',y')$, while the distance from BS to the $r$-th IRS element $d_{r}^{\rm{BR}}$ is fixed. Moreover, we set $f_{m}=f_{c}$ and $d_{r}^{\rm{RU'}}=d_{r}^{\rm{RU}}$, (28) can be re-expressed as
\begin{align}
g(f_{c},d_{r}^{\rm{RU}},\phi_{r}^{\rm{near}})=&\left\vert{\bf{b}}^T((1+\frac{f_{c}}{f_{c}})(d_{r}^{\rm{BR}}+d_{r}^{\rm{RU}}))\pmb{\theta}^{\rm{near}}\right\vert\nonumber\\
=&\left\vert \sum_{r=1}^{R}e^{-j\frac{2\pi}{\lambda_{c}}(1+\frac{f_{c}}{f_{c}})(d_{r}^{\rm{BR}}+d_{r}^{\rm{RU}})}e^{j\phi_{r}^{\rm{near}}} \right\vert \nonumber\\
=&\left\vert \sum_{r=1}^{R}e^{j[\phi_{r}^{\rm{near}}-\frac{4\pi}{\lambda_{c}}(d_{r}^{\rm{BR}}+d_{r}^{\rm{RU}})]} \right\vert.
\end{align}

When $g(f_{c},d_{r}^{\rm{RU}},\phi_{r}^{\rm{near}})$ reaches its maximum value, we have
\begin{eqnarray}
\phi_{r}^{\rm{near}}-\frac{4\pi}{\lambda_{c}}(d_{r}^{\rm{BR}}+d_{r}^{\rm{RU}})=0,
\end{eqnarray}
and
\begin{eqnarray}
\phi_{r}^{\rm{near}}=\frac{4\pi}{\lambda_{c}}(d_{r}^{\rm{BR}}+d_{r}^{\rm{RU}}).
\end{eqnarray}

Therefore, the optimal reflecting phase shift vector $\pmb{\theta}_{c}^{\rm{near}}$ can be expressed as
\begin{eqnarray}
\pmb{\theta}_{c}^{near}=[e^{j\frac{4\pi}{\lambda_{c}}(d_{1}^{\rm{BR}}+d_{1}^{\rm{RU}})},e^{j\frac{4\pi}{\lambda_{c}}(d_{2}^{\rm{BR}}+d_{2}^{\rm{RU}})},
\cdots,e^{j\frac{4\pi}{\lambda_{c}}(d_{R}^{\rm{BR}}+d_{R}^{\rm{RU}})}]^T.
\end{eqnarray}

Similar to (20), the  beam gain obtained with the reflecting phase shift vector $\pmb{\theta}_{c}^{near}$ at location $(x',y')$ and subcarrier frequency $f_{m}$ can be denoted as
\begin{align}
g(f_{m},d_{r}^{\rm{RU'}},\phi_{r}^{\rm{near}})=&\left\vert{\bf{b}}^T((1+\frac{f_{m}}{f_{c}})(d_{r}^{\rm{BR}}+d_{r}^{\rm{RU'}})){\pmb{\theta}_{c}^{near}}\right\vert \nonumber\\
=&\left\vert \sum_{r=1}^{R}e^{-j\frac{2\pi}{\lambda_{c}}(1+\frac{f_{m}}{f_{c}})(d_{r}^{\rm{BR}}+d_{r}^{\rm{RU'}})}e^{j\frac{4\pi}{\lambda_{c}}(d_{r}^{\rm{BR}}+d_{r}^{\rm{RU}})} \right\vert \nonumber\\
=&\left\vert \sum_{r=1}^{R}e^{j\frac{2\pi}{\lambda_{c}}[2(d_{r}^{\rm{BR}}+d_{r}^{\rm{RU}})-(1+\frac{f_{m}}{f_{c}})(d_{r}^{\rm{BR}}+d_{r}^{\rm{RU'}})]} \right\vert.
\end{align}

It is obvious that the maximum value of (33) can be obtained when $2(d_{r}^{\rm{BR}}+d_{r}^{\rm{RU}})-(1+\frac{f_{m}}{f_{c}})(d_{r}^{\rm{BR}}+d_{r}^{\rm{RU'}})=0$, and then the distance $d_{r}^{\rm{RU'}}$ between the $r$-th IRS element and location $(x',y')$ can be written as
\begin{eqnarray}
d_{r}^{\rm{RU'}}=\frac{2(d_{r}^{\rm{BR}}+d_{r}^{\rm{RU}})}{1+{f_{m}}/{f_{c}}}-d_{r}^{\rm{BR}}.
\end{eqnarray}

From (34), one can observe that the distance $d_{r}^{\rm{RU'}}$  can be approximated as $d_{r}^{\rm{RU'}}\thickapprox{d_{r}^{\rm{RU}}}$ over a given frequency band in narrowband systems due to ${f_{m}}/{f_{c}}\thickapprox1$. This means that the beams at all subcarriers will focus on the same location $(x,y)$ so that the high  beam gain of all subcarriers can be obtained at location $(x,y)$. On the contrary, one can observe that the beams at different subcarriers will be spread over different locations in wideband THz IRS systems, which is a new phenomenon referred to as the beam squint in the near-field.  We will show the above phenomenon in simulation results section.

When $(x',y')=(x,y)$, we have ${\bf{b}}^T((1+\frac{f_{m}}{f_{c}})(d_{r}^{\rm{BR}}+d_{r}^{\rm{RU'}}))={\bf{b}}^T((1+\frac{f_{m}}{f_{c}})(d_{r}^{\rm{BR}}+d_{r}^{\rm{RU}}))$. In this case, (35) can be expressed as
\begin{align}
g(f_{m},d_{r}^{\rm{RU}},\phi_{r}^{\rm{near}})=&\left\vert{\bf{b}}^T((1+\frac{f_{m}}{f_{c}})(d_{r}^{\rm{BR}}+d_{r}^{\rm{RU}})){\pmb{\theta}_{c}^{near}}\right\vert \nonumber\\
=&\left\vert \sum_{r=1}^{R}e^{j\frac{2\pi}{\lambda_{c}}[(1-\frac{f_{m}}{f_{c}})(d_{r}^{\rm{BR}}+d_{r}^{\rm{RU}})]} \right\vert.
\end{align}

Based on the above analysis, when $(x',y')=(x,y)$, the phase $e^{j\frac{2\pi}{\lambda_{c}}[(1-\frac{f_{m}}{f_{c}})(d_{r}^{\rm{BR}}+d_{r}^{\rm{RU}})]}$ will change with the IRS element $r$, which suffers the beam gain loss for several subcarriers. We will show the above phenomenon in simulation results section.
\section{Proposed Schemes for Overcoming Far-field and Near-Field Beam Squint Effects}
In Section \uppercase\expandafter{\romannumeral3}, we have analyzed the far-field and near-field beam squint effects, respectively. The results have shown exist serious beam gain loss due to the fact that the reflecting beams by IRS can not focus on the desired direction or location at all subcarriers. To address the above issue, inspired by the frequency-dependent characteristic of time delay line, we propose an effective scheme based on the DAM technique. Specifically, we apply the DAM to IRS so that the delay of the reflected signals by IRS can be adjusted, which can effectively mitigate the far-field and near-field beam squints effects by optimizing the reflection coefficients and delays of the IRS elements.

\subsection{The Proposed Scheme for Overcoming Far-Field Beam squint Effect}
Based on the DAM-based IRS structure, the frequency response of the $r$-th cascaded BS-IRS-user channel with DAM can be expressed as
\begin{eqnarray}
h_{r}^{\rm{D,far}}(f)=\alpha{e^{-j2\pi(r-1)\frac{d}{\lambda_{c}}(1+\frac{f}{f_{c}})\nu}}e^{-j2\pi{f}\tau_{r}^{\rm{D,far}}}e^{-j2\pi{f}\tau_{0}},
\end{eqnarray}
where ${\pmb{\tau}^{\rm{D,far}}}=[\tau_{1}^{\rm{D,far}},\tau_{2}^{\rm{D,far}},\cdots,\tau_{R}^{\rm{D,far}}]^T$ denotes time delays at IRS. We denote $g(f_{m},\nu,\phi_{r},\tau_{r}^{\rm{D,far}})=\left\vert{\bf{a}}^T((1+\frac{f_{m}}{f_{c}})\varphi)\pmb{\theta}_{c}{\pmb{\tau}^{\rm{D,far}}}\right\vert$ as the beam gain at subcarrier frequency $f_{m}$, which can be represented~as
\begin{align}
&g(f_{m},\nu_{0},\phi_{r},\tau_{r}^{\rm{D,far}})\nonumber\\
=&\left\vert{\bf{a}}^T((1+\frac{f_{m}}{f_{c}})\varphi_{0})\pmb{\theta}_{c}^{\rm{far}}{\pmb{\tau}^{\rm{D,far}}}\right\vert \nonumber\\
=&\left\vert \sum_{r=1}^{R}e^{-j2\pi(r-1)\frac{d}{\lambda_{c}}(1+\frac{f_{m}}{f_{c}})\nu_{0}}{e^{j2\pi(r-1)\nu_{0}}}e^{-j2\pi{f_{m}}\tau_{r}^{\rm{D,far}}}
\right\vert \nonumber\\
=&\left\vert \sum_{r=1}^{R}e^{j[2\pi(r-1)\nu_{0}-\pi(r-1)(1+\frac{f_{m}}{f_{c}})\nu_{0}-2\pi{f_{m}}\tau_{r}^{\rm{D,far}}]} \right\vert\nonumber\\
=&\left\vert \sum_{r=1}^{R}e^{j[\pi(r-1)\nu_{0}-\pi(r-1)\frac{f_{m}}{f_{c}}\nu_{0}-2\pi{f_{m}}\tau_{r}^{\rm{D}}]}\right\vert.
\end{align}
where $\tau_{r}^{\rm{D,far}}$ characterizes the time delay imposed by the $r$-th DAM. To maximize the beam gain, we have
\begin{align}
-2\pi{f_{m}}\tau_{r}^{\rm{D,far}}=&\pi(r-1)\frac{f_{m}}{f_{c}}\nu_{0}-\pi(r-1)\nu_{0}\nonumber\\
=&-\pi(r-1)\nu_{0}(1-\frac{f_{m}}{f_{c}}).
\end{align}

From (38), one can observe that $\tau_{r}^{\rm{D,far}}$ is decided by angle $\nu_{0}$ and subcarrier frequencies $f_{m}$. Next, we divide $\pi(r-1)\frac{f_{m}}{f_{c}}\nu_{0}-\pi(r-1)\nu_{0}$ into $\pi(r-1)\frac{f_{m}}{f_{c}}\nu_{0}$ and $-\pi(r-1)\nu_{0}$. Considering $\pi(r-1)\frac{f_{m}}{f_{c}}\nu_{0}$ is frequency-dependent, we set $\tau_{r}^{\rm{D}}$ as
\begin{eqnarray}
-2\pi{f_{m}}\tau_{r}^{\rm{D,far}}=\pi(r-1)\frac{f_{m}}{f_{c}}\nu_{0},
\end{eqnarray}
then, $\tau_{r}^{\rm{D,far}}$ can be written as
\begin{eqnarray}
\tau_{r}^{\rm{D,far}}=-(r-1){\nu_{0}}/{2f_{c}}.
\end{eqnarray}

From (40), one can see that $\tau_{r}^{\rm{D,far}}$ is a function of $r$, $f_{c}$ and $\nu_{0}$, which can be applied to all subcarriers due to its frequency independent nature.

Next, the second part $-\pi(r-1)\nu_{0}$ can be realized by adding an extra phase shift at IRS. Therefore, the optimal reflecting coefficient $\pmb{\theta}^{\rm{far}}_{c}$ can be written as
\begin{eqnarray}
\pmb{\theta}^{\rm{far}}_{c}=[1,e^{j2\pi\nu_{0}}e^{-j\pi\nu_{0}},\cdots,e^{j2\pi(R-1)\nu_{0}}e^{-j\pi(R-1)\nu_{0}}]^T.
\end{eqnarray}

Consequently, the phase shift of the $r$-th IRS element is
\begin{eqnarray}
\phi^{\rm{far}}_{r,c}=\pi(r-1)\nu_{0}.
\end{eqnarray}

Considering $\tau_{r}^{\rm{D}}$ is larger than $0$, the time delay of the $r$-th IRS DAM can be rewritten as
\begin{equation}
\tau_{r}^{\rm{D,far}}=\left\{
\begin{aligned}
-(r-1)\frac{\nu_{0}}{2f_{c}}       &      & {\nu_{0}      \leqslant      0},\\
(R-1)\frac{\nu_{0}}{2f_{c}}-(r-1)\frac{\nu_{0}}{2f_{c}}       &      & {\nu_{0}      >      0}.
\end{aligned} \right.
\end{equation}

Based on the obtained IRS reflecting coefficients and time delays, the beam gain of the $m$-th subcarrier can be expressed~as
\begin{align}
&g(f_{m},\nu,\phi_{r}^{\rm{far}},\tau_{r}^{\rm{D,far}}) \nonumber\\
=&\left\vert{\bf{a}}^T((1+\frac{f_{m}}{f_{c}})\varphi)\pmb{\theta}^{\rm{far}}_{c}{\pmb{\tau}^{\rm{D,far}}}\right\vert \nonumber\\
=&\left\vert \sum_{r=1}^{R}e^{-j2\pi(r-1)\frac{d}{\lambda_{c}}(1+\frac{f_{m}}{f_{c}})\nu}e^{j\pi(r-1)\nu_{0}}e^{-j2\pi{f_{m}}\tau_{r}^{\rm{D,far}}}
\right\vert \nonumber\\
=&\left\vert \sum_{r=1}^{R}e^{-j\pi(r-1)(1+\frac{f_{m}}{f_{c}})\nu}e^{j\pi(r-1)(1+\frac{f_{m}}{f_{c}})\nu_{0}} \right\vert \nonumber\\
=&\left\vert\sum_{r=1}^{R}e^{j\pi(r-1)[(1+\frac{f_{m}}{f_{c}})(\nu_{0}-\nu)]} \right\vert.
\end{align}

It is obvious that (44) can always reach its maximum value for any subcarrier as long as $\nu_{0}=\nu$. This means that the beam squint effects can be eliminated by optimizing the reflecting coefficient and time delays to satisfy $\nu_{0}=\nu$.
\subsection{The Proposed Scheme for Overcoming Near-Field Beam squint Effect}
Similarly, the near-field frequency response of the $r$-th cascaded BS-IRS-user channel at location $(x,y)$ can be expressed~as
\begin{align}
h_{r}^{\rm{D,near}}(f)={\alpha}e^{-j\frac{2\pi}{\lambda_{c}}(1+\frac{f}{f_{c}})(d_{r}^{\rm{BR}}+d_{r}^{\rm{RU}})}
e^{-j2\pi{f}\tau_{r}^{\rm{D,near}}},
\end{align}
where ${\pmb{\tau}^{\rm{D,near}}}=[\tau_{1}^{\rm{D,near}},\tau_{2}^{\rm{D,near}},\cdots,\tau_{R}^{\rm{D,near}}]^T$ denotes the time delays at IRS, and $\tau_{r}^{\rm{D,near}}$ characterizes the time delay imposed by the $r$-th IRS element. Additionally, the beam gain $g(f_{m},d_{r}^{\rm{RU}},\phi_{r}^{\rm{near}})=\left\vert{\bf{b}}^T((1+\frac{f_{m}}{f_{c}})(d_{r}^{\rm{BR}}+d_{r}^{\rm{RU}}))\pmb{\theta}_{c}^{\rm{near}}
e^{-j2\pi{f_{m}}\tau_{r}^{\rm{D,near}}}\right\vert$ at location $(x,y)$ can be rewritten as
\begin{align}
&g(f_{m},d_{r}^{\rm{RU}},\phi_{r}^{\rm{near}})\nonumber\\
=&\left\vert{\bf{b}}^T((1+\frac{f_{m}}{f_{c}})(d_{r}^{\rm{BR}}+d_{r}^{\rm{RU}}))\pmb{\theta}_{c}^{\rm{near}}
e^{-j2\pi{f_{m}}\tau_{r}^{\rm{D,near}}}\right\vert\nonumber\\
=&\left\vert \sum_{r=1}^{R}e^{-j\frac{2\pi}{\lambda_{c}}(1+\frac{f_{m}}{f_{c}})(d_{r}^{\rm{BR}}+d_{r}^{\rm{RU}})}
e^{j\phi_{r,c}^{\rm{near}}}e^{-j2\pi{f_{m}}\tau_{r}^{\rm{D,near}}}\right\vert \nonumber\\
=&\left\vert \sum_{r=1}^{R}e^{j\frac{2\pi}{\lambda_{c}}[(d_{r}^{\rm{BR}}+d_{r}^{\rm{RU}})(1-\frac{f_{m}}{f_{c}})]
-j2\pi{f_{m}}\tau_{r}^{\rm{D,near}}}\right\vert.
\end{align}

Similarly, to obtain the maximum beam gain, we have
\begin{eqnarray}
\frac{2\pi}{\lambda_{c}}[(d_{r}^{\rm{BR}}+d_{r}^{\rm{RU}})(1-\frac{f_{m}}{f_{c}})]
-2\pi{f_{m}}\tau_{r}^{\rm{D,near}}=0.
\end{eqnarray}

According to (47), the phase adjusted by DAM can be written as
\begin{eqnarray}
-2\pi{f_{m}}\tau_{r}^{\rm{D,near}}=\frac{2\pi}{\lambda_{c}}(d_{r}^{\rm{BR}}+d_{r}^{\rm{RU}})\frac{f_{m}}{f_{c}}
-\frac{2\pi}{\lambda_{c}}(d_{r}^{\rm{BR}}+d_{r}^{\rm{RU}}).
\end{eqnarray}

Similar to the previous analysis, we divide $-2\pi{f_{m}}\tau_{r}^{\rm{D,near}}$ into $\frac{2\pi}{\lambda_{c}}(d_{r}^{\rm{BR}}+d_{r}^{\rm{RU}})\frac{f_{m}}{f_{c}}$ and $-\frac{2\pi}{\lambda_{c}}(d_{r}^{\rm{BR}}+d_{r}^{\rm{RU}})$. It is obvious that the former is frequency-dependent and can be achieved by setting $\tau_{r}^{\rm{D,near}}$ as
\begin{eqnarray}
-2\pi{f_{m}}\tau_{r}^{\rm{D,near}}=\frac{2\pi}{\lambda_{c}}(d_{r}^{\rm{BR}}+d_{r}^{\rm{RU}})\frac{f_{m}}{f_{c}}.
\end{eqnarray}
\begin{figure*}[t]
	\begin{center}
		\includegraphics[scale = 0.6]{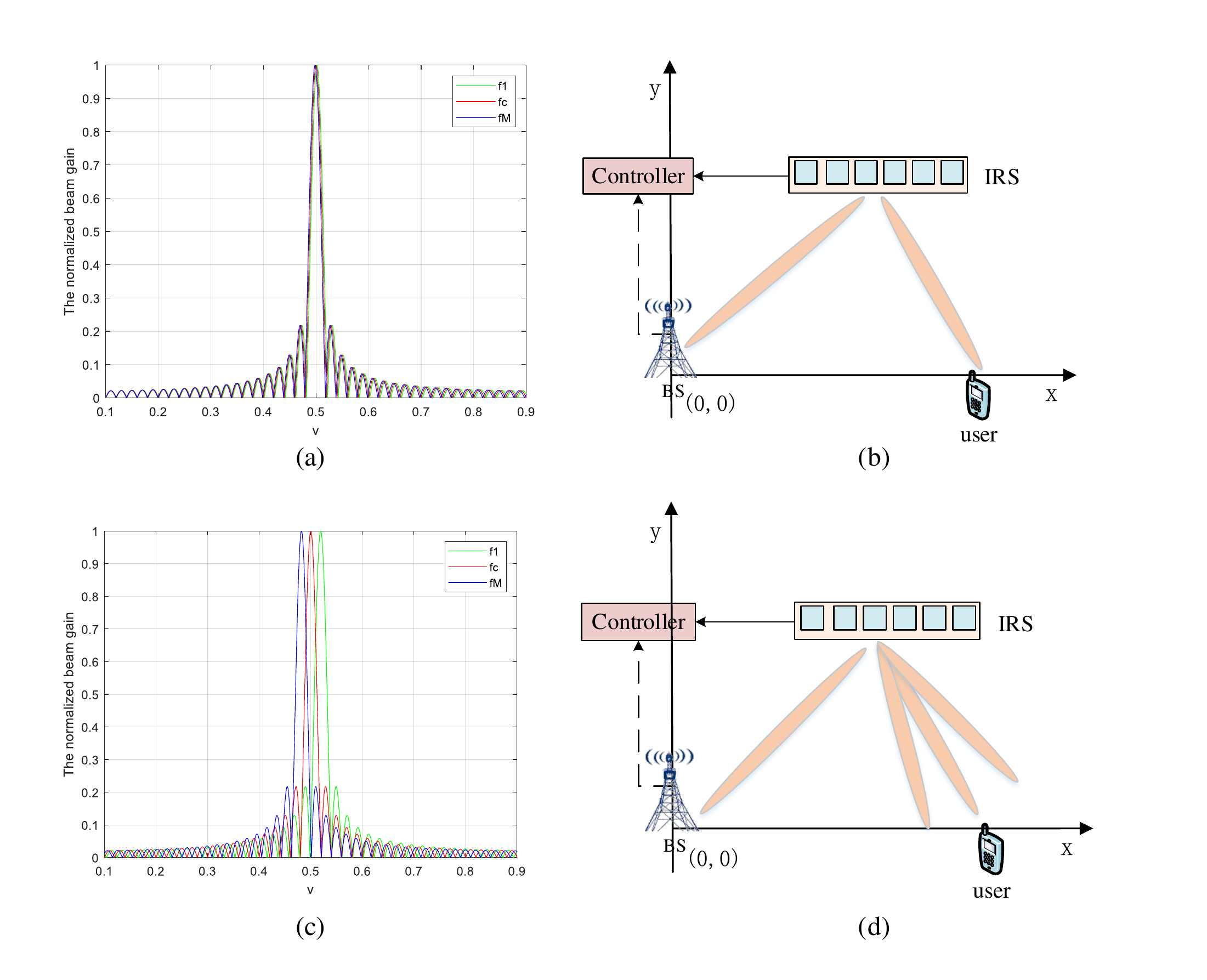}
		\caption{(a) The normalized beam gain versus the angle in narrowband IRS systems. (b) The beam model in narrowband IRS systems. (c) The normalized beam gain versus the angle in wideband systems. (d) The far-field beam squint model in THz IRS systems.}
		\label{figure2}
	\end{center}
\end{figure*}
Thus, $\tau_{r}^{\rm{D,near}}$ can be expressed as
\begin{eqnarray}
\tau_{r}^{\rm{D,near}}=-\frac{d_{r}^{\rm{BR}}+d_{r}^{\rm{RU}}}{c}.
\end{eqnarray}

From (50), one can observe that the $r$-th IRS time delay in the near-field is only relevant to the distance from BS to the $r$-th IRS and from the $r$-th IRS to the location $(x,y)$ and the speed of light $c$, which is straightforward to implement for all subcarriers.

Next, the second part $-\frac{2\pi}{\lambda_{c}}(d_{r}^{\rm{BR}}+d_{r}^{\rm{RU}})$ can be obtained by DAM, namely setting an extra phase shift to the IRS element. Thus, the phase shifts provided by IRS can be written as
\begin{align}
\pmb{\theta}_{c}^{near}=&[e^{j\frac{4\pi}{\lambda_{c}}(d_{1}^{\rm{BR}}+d_{1}^{\rm{RU}})}e^{-j\frac{2\pi}{\lambda_{c}}(d_{1}^{\rm{BR}}+d_{1}^{\rm{RU}})},
e^{j\frac{4\pi}{\lambda_{c}}(d_{2}^{\rm{BR}}+d_{2}^{\rm{RU}})}e^{-j\frac{2\pi}{\lambda_{c}}(d_{2}^{\rm{BR}}+d_{2}^{\rm{RU}})},\nonumber\\
&\cdots,e^{j\frac{4\pi}{\lambda_{c}}(d_{R}^{\rm{BR}}+d_{R}^{\rm{RU}})}e^{-j\frac{2\pi}{\lambda_{c}}(d_{R}^{\rm{BR}}+d_{R}^{\rm{RU}})}]^T.
\end{align}

The phase shift of the $r$-th IRS element is
\begin{eqnarray}
\phi_{r}^{\rm{near}}=\frac{2\pi}{\lambda_{c}}(d_{r}^{\rm{BR}}+d_{r}^{\rm{RU}}).
\end{eqnarray}


However, one can observe that the time delay $-\frac{d_{r}^{\rm{BR}}+d_{r}^{\rm{RU}}}{c}$ is less than 0, we can add a common delay $T$ to satisfy $min\{\tau_{r}^{\rm{D,near}}\}\geqslant0$. Finally, the optimal time delay of the $r$-th IRS can be represented as
\begin{eqnarray}
\tau_{r}^{\rm{D,near*}}=T-\frac{d_{r}^{\rm{BR}}+d_{r}^{\rm{RU}}}{c}.
\end{eqnarray}

Based on the above analysis, we investigate the beam gain at location $(x',y')$ based on the near-field. By substituting (52) and (53) into $g(f_{m},d_{r}^{\rm{RU'}},\phi_{r}^{\rm{near}},\tau_{r}^{\rm{D,near*}})=\left\vert{\bf{b}}^T((1+\frac{f_{m}}{f_{c}})(d_{r}^{\rm{BR}}+d_{r}^{\rm{RU'}}))\pmb{\theta}_{c}^{\rm{near}}
e^{-j2\pi{f_{m}}\tau_{r}^{\rm{D,near*}}}\right\vert$ over a given frequency band, the beam gain can be expressed as

\begin{align}
&g(f_{m},d_{r}^{\rm{RU'}},\phi_{r}^{\rm{near}},\tau_{r}^{\rm{D,near*}})\nonumber\\
=&\left\vert{\bf{b}}^T((1+\frac{f_{m}}{f_{c}})(d_{r}^{\rm{BR}}+d_{r}^{\rm{RU'}}))\pmb{\theta}_{c}^{\rm{near}}
e^{-j2\pi{f_{m}}\tau_{r}^{\rm{D,near*}}}\right\vert\nonumber\\
=&\left\vert \sum_{r=1}^{R}e^{-j\frac{2\pi}{\lambda_{c}}(1+\frac{f_{m}}{f_{c}})(d_{r}^{\rm{BR}}+d_{r}^{\rm{RU'}})}
e^{j\frac{2\pi}{\lambda_{c}}(d_{r}^{\rm{BR}}+d_{r}^{\rm{RU}})}e^{-j2\pi{f_{m}}(T-\frac{d_{r}^{\rm{BR}}+d_{r}^{\rm{RU}}}{c})}\right\vert \nonumber\\
=&\left\vert \sum_{r=1}^{R}e^{-j\frac{2\pi}{\lambda_{c}}(1+\frac{f_{m}}{f_{c}})(d_{r}^{\rm{BR}}+d_{r}^{\rm{RU'}})}
e^{j\frac{2\pi}{\lambda_{c}}(1+\frac{f_{m}}{f_{c}})(d_{r}^{\rm{BR}}+d_{r}^{\rm{RU}})} \right\vert \nonumber\\
=&\left\vert\sum_{r=1}^{R}e^{j\frac{2\pi}{\lambda_{c}}(1+\frac{f_{m}}{f_{c}})[(d_{r}^{\rm{BR}}+d_{r}^{\rm{RU}})-
(d_{r}^{\rm{BR}}+d_{r}^{\rm{RU'}})]} \right\vert \nonumber\\
=&\left\vert\sum_{r=1}^{R}e^{j\frac{2\pi}{\lambda_{c}}(1+\frac{f_{m}}{f_{c}})(d_{r}^{\rm{RU}}-d_{r}^{\rm{RU'}})} \right\vert.
\end{align}

From (54), it is apparent that the maximum beam gain can be obtained for any subcarrier at location $(x',y')$, namely $d_{r}^{\rm{RU}}=d_{r}^{\rm{RU'}}$. This means that the beam squint effect can also be elimated by optimizing the reflecting coefficients and time delays of the IRS. We will present the performance analysis in Section V.

\section{Numerical Results}
In this section, numerical results are presented to illustrate the beam squint effects and the effectiveness of our proposed DAM approaches in the IRS-aided THz systems, respectively. Unless otherwise stated, we set $L_{1}=L_{2}=1$, $M=128$, $f_{c}=200$ GHz and $B=6$ GHz.
\subsection{Far-Field Beam squint Effects and the Proposed Solution}
\begin{figure}[t]
	\begin{center}
		\includegraphics[scale = 0.6]{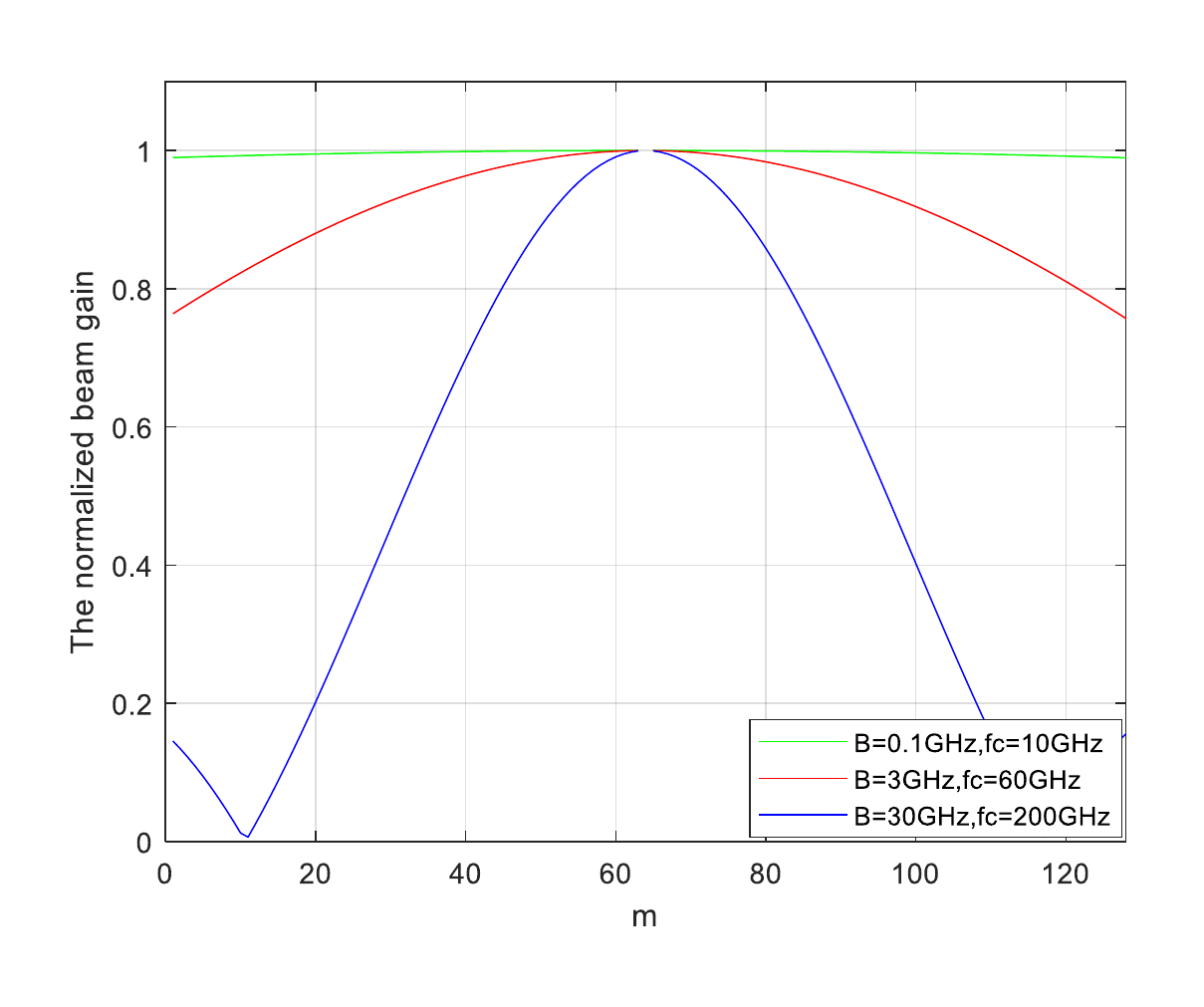}
		\caption{The normalized beam gain versus subcarrier under different bandwidths.}
		\label{figure3}
	\end{center}
\end{figure}

To better account for the far-field beam squint effect, Fig. 2 shows the normalized beam gain at different subcarriers against an arbitrary angle $\upsilon$ with a fixed distance from BS to IRS in narrowband systems and wideband systems, respectively. We denote $f_{1}$, $f_{c}$ and $f_{M}$ as the lowest, the central and the highest subcarrier frequency, respectively. Obviously, Fig. 2(a) illustrates that the beam from IRS to the user at subcarrier frequency $f_{1}$ and $f_{M}$ can achieve the highest gain in narrow systems. In other words, beams at all subcarriers can point at the same direction as shown in Fig. 2(b).

On the contrary, from Fig. 2(c), we can observe that the far-field beam squint effect makes that the beams at different subcarriers steer toward different directions. We also plot the beam squint model as shown in Fig. 2(d), where the user is located at the far-field region and the beams reflected by IRS fail to focus on the user's direction, leading to the serious gain loss over the given bandwidth as discussed in Section
\uppercase\expandafter{\romannumeral3}.

\begin{figure}[t]
	\begin{center}
		\includegraphics[scale = 0.6]{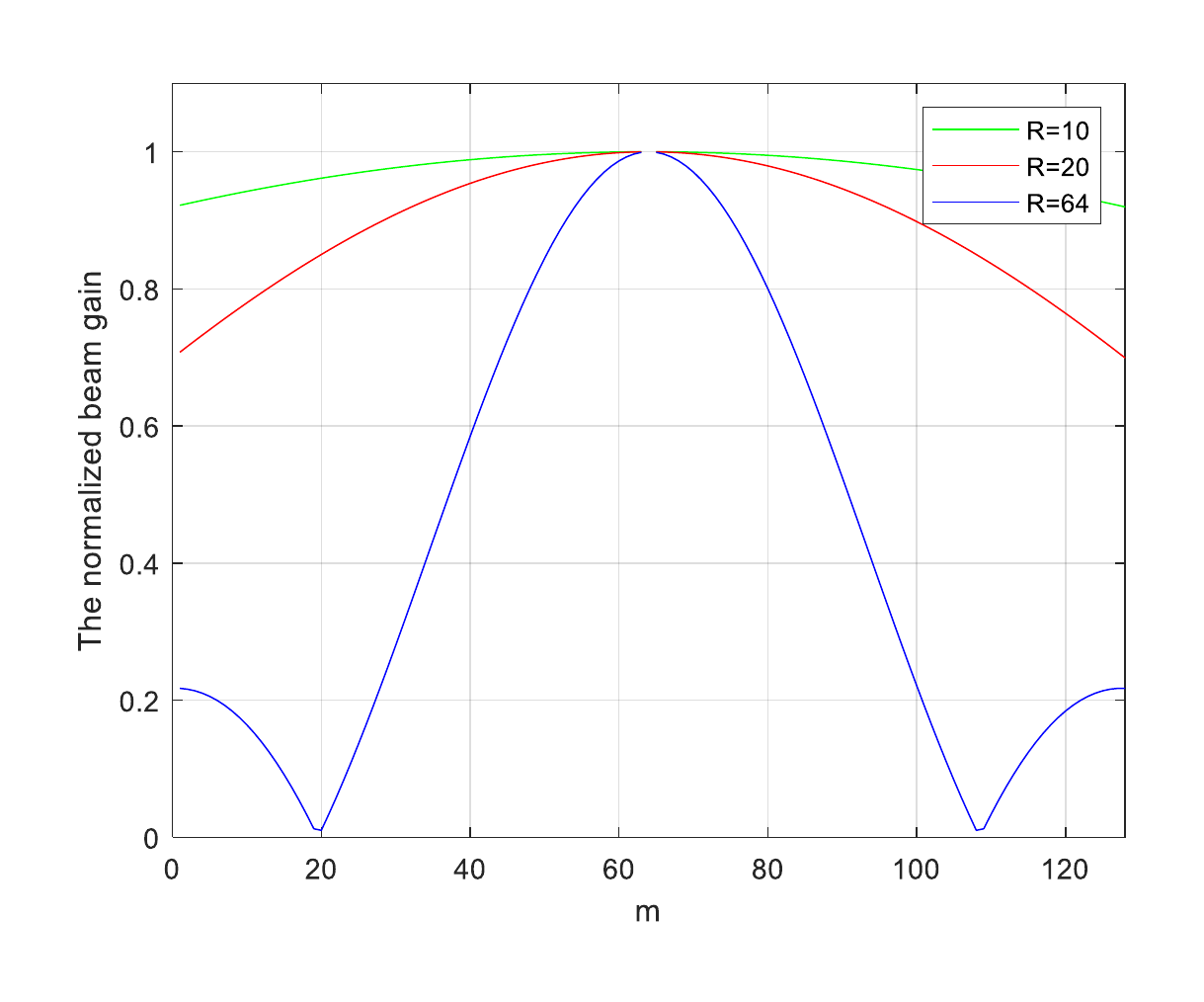}
		\caption{The normalized beam gain versus the subcarriers under different IRS elements.}
		\label{figure4}
	\end{center}
\end{figure}
\begin{figure}[t]
	\begin{center}
		\includegraphics[scale = 0.6]{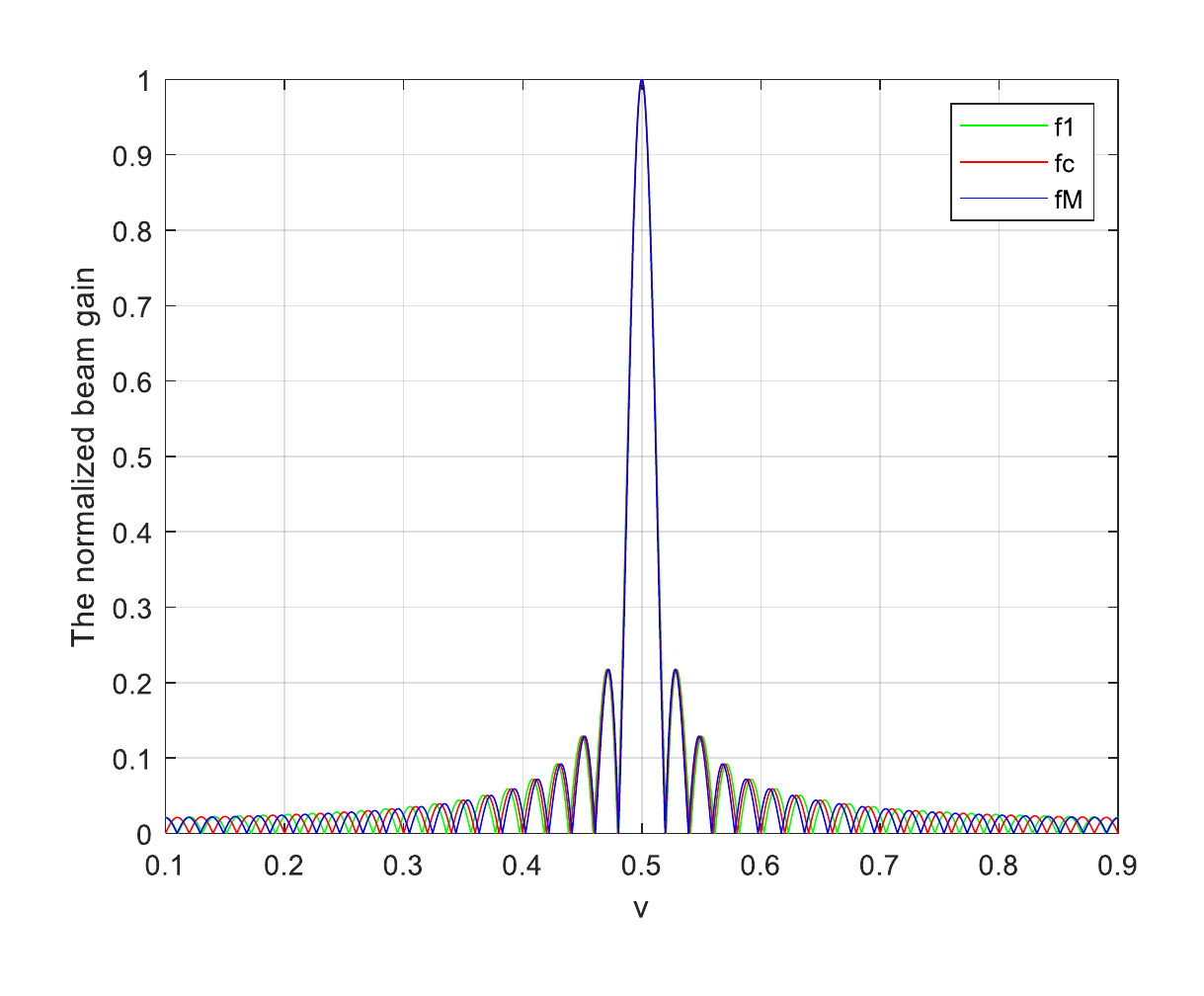}
		\caption{The normalized beam gain and beam pattern against the angle under the proposed scheme.}
		\label{figure5}
	\end{center}
\end{figure}
\begin{figure*}[t]
	\begin{center}
		\includegraphics[scale = 0.7]{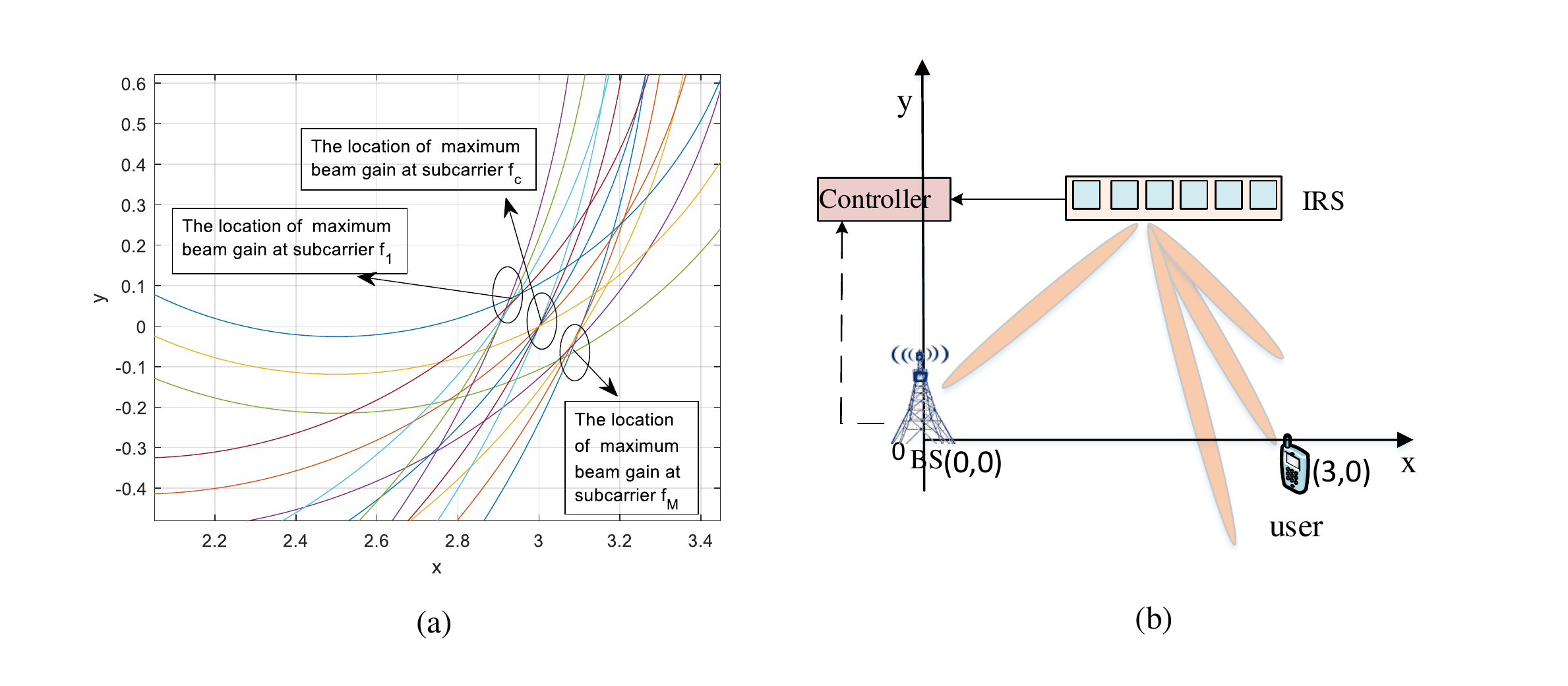}
		\caption{(a) The location of  maximum beam gain at subcarriers $f_{1}$,$f_{c}$ and $f_{M}$, respectively. (b) The near-field beam squint model in THz IRS communications.}
		\label{figure6}
	\end{center}
\end{figure*}
\begin{figure}[t]
	\begin{center}
		\includegraphics[scale = 0.6]{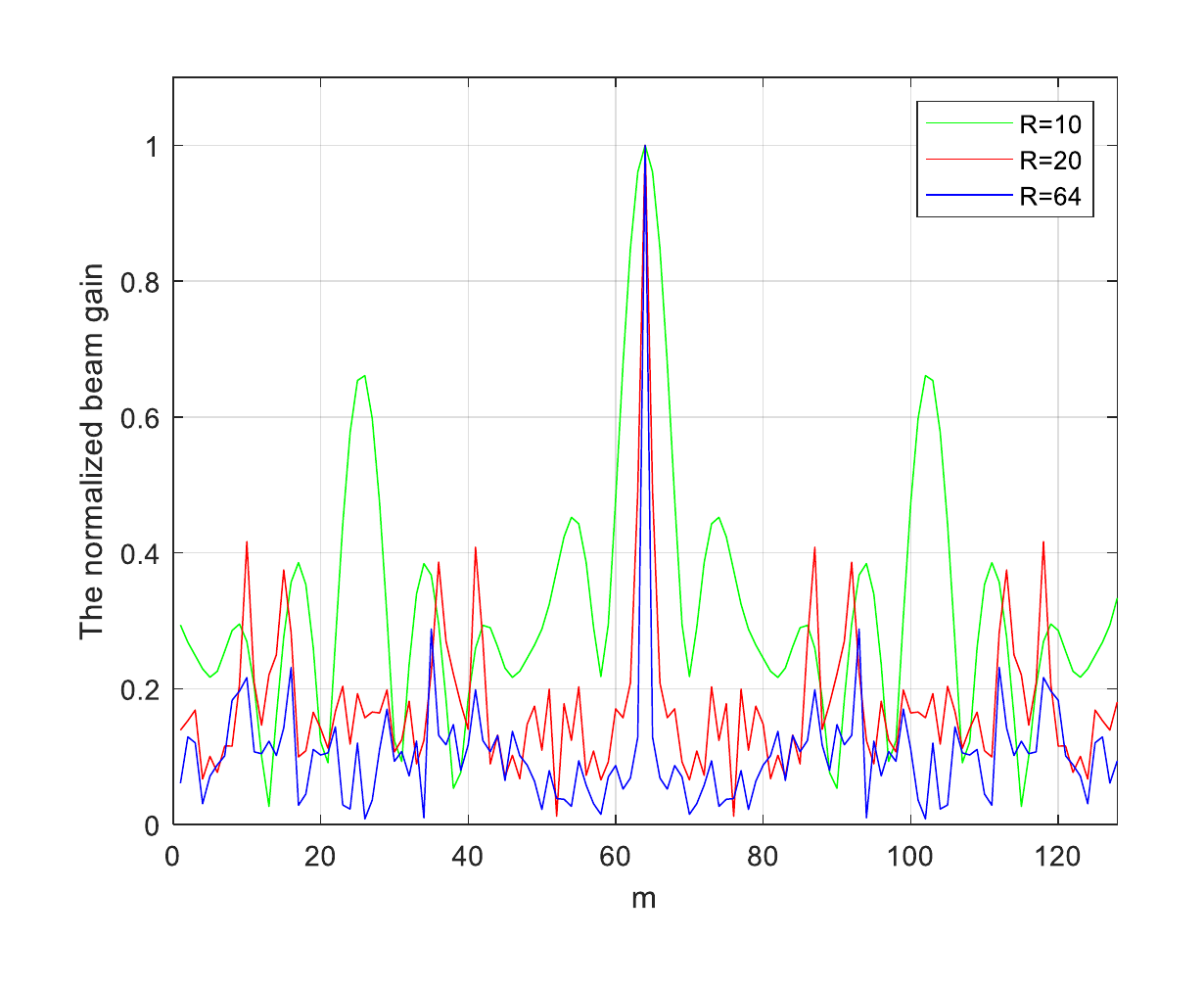}
		\caption{The near-field beam squint effect under different IRS elements.}
		\label{figure7}
	\end{center}
\end{figure}

According to (23), we further evaluate the beam squint effect under different frequencies, bandwidth and IRS elements in Figs. 3 and 4. It can be observed from Fig. 3 that the maximum gain can be achieved almost across all subcarriers when the bandwidth and frequency are relatively small. Whereas, with the bandwidth and frequency increases, the beam gain loss rapidly increases at different subcarriers. For example, 70$\%$ subcarriers obtain only its 20$\%$ beam gain when $B=30$ GHz, $f_{c}=200$ GHz. This means that as the bandwidth $B$ increases, the beam squint effect becomes more pronounced. In addition, Fig.~4 shows the normalized beam gain versus subcarries with different IRS elements. Similarly, one can observe that with the IRS elements increase, the beam squint effect becomes more obvious. For example, the united beam gain at all subcarries almost the same when $R=10$. However, when $R=64$, we find that only 23$\%$ subcarriers can obtain a relatively high beam gain.
Fig. 5 illustrates the normalized beam gain and beam pattern against the angle under the proposed scheme. Obviously, we can find that the beams at other subcarriers are focused on the spatial direction $\nu_{0}$ and the generated beams can achieve nearly 98$\%$ gain at the desired direction, which show the effectiveness of our proposed scheme.
\subsection{Near-Field Beam Squint Effect and The Proposed Solution}
In the near-field case, we set the locations of BS, user and the first IRS element as $(0,0)$, $(3,0)$ and $(1,1)$, respectively. To better explain near-field beam squint effect, we estimate and draw the locations of beams at subcarrier frequency $f_{1}$, $f_{c}$ and $f_{M}$.



Fig. 6 plots the near-field beam squint phenomenon in THz IRS communications. Similarly, we first set IRS phase shifters to place the maximum beam gain at subcarrier $f_{c}$ on the user. Then, we plot the locations of maximum beam gains at subcarriers $f_{1}$ and $f_{M}$, as shown in Fig.~6(a). It is clear that the near-field beam squint causes the maximum beam gains for different subcarriers to occur different locations, which is different from the far-field beam squint. To show the near-field beam squint more clearly, we also plot the beam transmission pattern in Fig.~6(b). In addition, we plot the near-field beam squint effect under different IRS elements in Fig.~7 by showing the normalized beam gains at different subcarriers. On can observe the beam gain loss for most of subcarries. Furthermore, the beam gain loss increases as the IRS elements increase.


Fig. 8 shows the beam pattern after applying proposed scheme. It can be clearly seen that the beams at all subcarriers nealy focus on the same location as shown in Fig.~8(a), which show the effectiveness of the proposed scheme. Meanwhile, to clearly demonstrate that the beam squint effect has been eliminated, we also plot the beam model in Fig.~8(b). Thus, we can conclude from Fig. 8 that the near-field beam squint effect also can be overcome by our proposed scheme.
\begin{figure*}[t]
	\begin{center}
		\includegraphics[scale = 0.7]{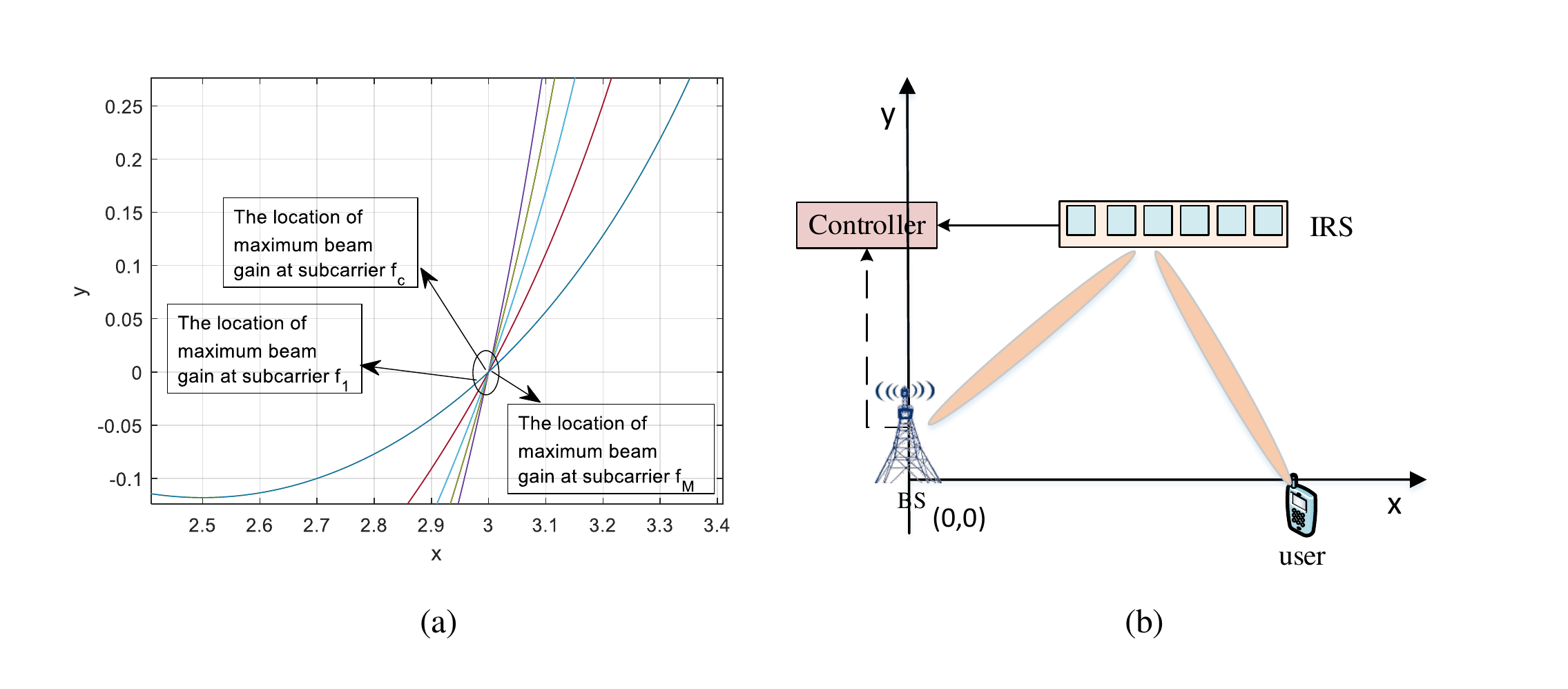}
		\caption{(a) The location of maximum beam gain at subcarriers $f_{1}$,$f_{c}$ and $f_{M}$, respectively. (b)The near-field beam squint model after applying proposed scheme.}
		\label{figure8}
	\end{center}
\end{figure*}
\section{Conclusion}
In this paper, we first pointed out the far-field and near-field beam squint effects in an IRS-aided wideband THz communications. Specially, the far-field beam squint makes that beams at different subcarriers squint toward different directions, while the near-field beam squint spreads beams at different subcarriers point over different locations. Next, we proposed a scheme that applying DAM to IRS for overcoming the far-field and near-field beam squint effects, through joint optimization of the time delays and reflection phase shifts of IRS elements. Simulations results showed that our proposed schemes are effective to deal with beam squint effects in THz IRS communications. The far-field and near-field beam squint issues in multi-user MIMO systems will be considered in our future work.


\begin{thebibliography}{99}
\bibitem{1a}
S. Bani Melhem and H. Tabassum, ``User pairing and outage analysis in multi-carrier NOMA-THz networks,"
 \textit{IEEE Trans. Veh. Tech.}, doi: 10.1109/TVT.2022.3146352.
\bibitem{2a}
H. Sarieddeen, M. Alouini, and T. Y. Al-Naffouri, ``Terahertz-band ultra-massive spatial modulation MIMO,"
\textit{IEEE J. Sel. Areas Commun.},
vol. 37, no. 9, pp. 2040-2052, Sep. 2019.
\bibitem{3a}
World Radiocommunication Conference 2019 (WRC-19): Provisional Final Acts,
\textit{ITU Publications}, Sharm El-Sheikh, Egypt, Oct. 2019.
\bibitem{4a}
B. Zhai, A. Tang, C. Peng, and X. Wang, ``SS-OFDMA: Spatial-spread orthogonal frequency division multiple access for terahertz networks,"
\textit{IEEE J. Sel. Areas Commun.}, vol. 39, no. 6, pp. 1678-1692, Jun. 2021.
\bibitem{5a}
Q. Wu and R. Zhang, ``Intelligent reflecting surface enhanced wireless network via joint active and passive beamforming,"
\textit{IEEE Trans. Wireless Commun.}, vol. 18, no. 11, pp. 5394-5409, Nov. 2019.
\bibitem{6a}
W. Hao, G. Sun, M. Zeng, Z. Chu, Z. Zhu, O. A. Dobre, and
P. Xiao, ``Robust design for intelligent reflecting surface-assisted MIMO-OFDMA terahertz loT networks,"
\textit{IEEE Internet Things J.}, vol. 8, no. 16, pp. 13052-13064, Aug. 2021.
\bibitem{7a}
B. Lyu, P. Ramezani, D. T. Hoang, and A. Jamalipour, ``IRS-assisted downlink and uplink NOMA in wireless powered communication networks,"
\textit{IEEE Trans. Veh. Tech.}, vol. 71, no. 1, pp. 1083-1088, Jan. 2022.
\bibitem{8}
P. Ramezani and A. Jamalipour, ``Backscatter-assisted wireless powered communication networks empowered by intelligent reflecting surface,"
\textit{IEEE Trans. Veh. Tech.}, vol. 70, no. 11, pp. 11908-11922, Nov. 2021.
\bibitem{8a}
 M. Cai, K. Gao, D. Nie, B. Hochwald, J. N. Laneman, H. Huang, and K. Liu, ``Effect of wideband beam squint on codebook design in phased array wireless systems,"
\textit{in Proc. IEEE GLOBECOM 2016}, pp. 1-6, Dec. 2016.
\bibitem{9a}
Y. Chen, D. Chen, and T. Jiang, ``Beam-squint mitigating in reconfigurable intelligent surface aided wideband mmWave communications,"
\textit{2021 IEEE Wireless Commun. Netw. Conf. (WCNC)}, pp. 1-6, 2021.
\bibitem{10a}
Q. Wan, J. Fang, Z. Chen, and H. Li, ``Hybrid precoding and combining for millimeter wave/sub-THz MIMO-OFDM systems with beam squint effects,"
\textit{IEEE Trans. Veh. Tech.}, vol. 70, no. 8, pp. 8314-8319, Aug. 2021.
\bibitem{11a}
J. Tan and L. Dai, ``Delay-phase precoding for THz massive MIMO with beam squint,"
\textit{in Proc. IEEE GLOBECOM}, pp. 1-6, 2019.
\bibitem{12a}
F. Gao, B. Wang, C. Xing, and et al., ``Wideband beamforming for hybrid massive MIMO terahertz communications,"
\textit{IEEE J. Sel. Areas Commun.}, vol. 39, no. 6, pp. 1725-1740, Jun. 2021.
\bibitem{13a}
B. Zhai, Y. Zhu, A. Tang, and X. Wang, ``THzPrism: Frequency-based beam spreading for terahertz communication systems,"
\textit{IEEE Wireless Commun. Lett.}, vol. 9, no. 6, pp. 897-900, Jun. 2020.
\bibitem{14a}
Z. Zhou, X. Gao, J. Fang, and Z. Chen, ``Spherical wave channel and analysis for large linear array in LoS conditions,"
\textit{in Proc. IEEE Globecom Workshops 2015}, pp. 1-6, Dec. 2015.
\bibitem{15a}
K. T. Selvan and R. Janaswamy, ``Fraunhofer and fresnel distances: Unified derivation for aperture antennas,"
\textit{IEEE Antennas Propag. Mag.}, vol. 59, no. 4, pp. 12-15, Aug. 2017.
\bibitem{16a}
M. Cui and L. Dai, ``Channel estimation for extremely large-scale MIMO: far-field or near-field?"
\textit{IEEE Trans. Commun.}, doi: 10.1109/TCOMM.2022.3146400.
\bibitem{17a}
M. Cui, et al., ``Near-field wideband beamforming for extremely large antenna array,"
\textit{arXiv preprint arXiv:2109.10054v1}, Sep. 2021.
\bibitem{18a}
Y. Zhang, J. Zhang, M. Di Renzo, H. Xiao, and B. Ai, ``Reconfigurable intelligent surfaces with outdated channel state information: Centralized vs. distributed deployments,"
\textit{IEEE Trans. Commun.}, doi: 10.1109/TCOMM.2022.3146344.
\bibitem{19a}
 J. An, C. Xu, D. W. K. Ng, C. Yuen, L. Gan, and L. Hanzo, ``Reconfigurable intelligent surface-enhanced OFDM communications via delay adjustable metasurface,"
\textit{arXiv preprint arXiv:2110.09291}, Oct. 2021.
\bibitem{20a}
S. Ma, W. Shen, J. An, and L. Hanzo, ``Wideband channel estimation for IRS-aided systems in the face of beam squint,"
\textit{IEEE Trans. Wireless Commun.}, vol. 20, no. 10, pp. 6240-6253, Oct. 2021.
\bibitem{21a}
M. Jian, F. Gao, Z. Tian, S. Jin, and S. Ma, ``Angle-domain aided UL/DL channel estimation for wideband mmWave massive MIMO systems with beam squint,"
\textit{IEEE Trans. Wireless Commun.}, vol. 18, no. 7, pp. 3515-3527, Jul. 2019.




\end{thebibliography}
\end{document}